\documentclass[pdflatex,sn-nature,centre]{sn-jnl}

\usepackage{graphicx}%
\usepackage{multirow}%
\usepackage{amsmath,amssymb,amsfonts}%
\usepackage{amsthm}%
\usepackage{mathrsfs}%
\usepackage[title]{appendix}%
\usepackage{xcolor}%
\usepackage{textcomp}%
\usepackage{manyfoot}%
\usepackage{booktabs}%
\usepackage{algorithm}%
\usepackage{algorithmicx}%
\usepackage{algpseudocode}%
\usepackage{listings}%
\usepackage{lmodern}

\begin{document}

\title[Article Title]{Cold-atom comagnetometry via {optical} control of spin states}


\author[1]{\fnm{J.-L.} \sur{Zhang}}\email{jielin@mail.ustc.edu.cn}
\equalcont{These authors contributed equally to this work.}

\author[1]{\fnm{W.-T.} \sur{Luo}}\email{luowt@mail.ustc.edu.cn}
\equalcont{These authors contributed equally to this work.}

\author[1]{\fnm{Y. A.} \sur{Yang}}\email{yyustcer@mail.ustc.edu.cn}

\author[1]{\fnm{Y.-Q.} \sur{Wang}}\email{wyq0705@mail.ustc.edu.cn}

\author*[2]{\fnm{T.} \sur{Xia}}\email{txia@hfnl.cn}

\author*[1,2]{\fnm{Z.-T.} \sur{Lu}}\email{ztlu@ustc.edu.cn}

\affil[1]{\orgdiv{Hefei National Research Center for Physical Sciences at the Microscale, School of Physical Sciences}, \orgname{University of Science and Technology of China}, \orgaddress{\city{Hefei}, \state{Anhui}, \postcode{230026}, \country{China}}}

\affil[2]{\orgdiv{Hefei National Laboratory}, \orgname{University of Science and Technology of China}, \orgaddress{\city{Hefei}, \state{Anhui}, \postcode{230088}, \country{China}}}


\abstract{
Atomic spin-based comagnetometers are powerful tools for precision sensing and tests of fundamental physics. Compared with the widely used gas-cell comagnetometer systems, cold-atom systems offer access to much shorter distance scales and allow implementation of {optical} quantum control techniques. However, in order to realize long spin coherence times with cold atoms, it is necessary to employ diamagnetic atoms and overcome decoherence induced by light shifts. Here we demonstrate a cold-atom comagnetometer based on the nuclear spins of $^{171}$Yb (spin-1/2) and $^{173}$Yb (spin-5/2), jointly trapped in an optical lattice. Vector light shifts are suppressed by enforcing linear polarization of the lattice, while tensor shifts in $^{173}$Yb are suppressed via the use of a Schr\"{o}dinger cat state. This enables simultaneous Ramsey interferometry on both isotopes with a spin coherence time of 60 s. We achieve a magnetic noise suppression factor exceeding $3\times10^4$, and determine the ratio of nuclear magnetic moments to 4 ppm precision. Our results establish a new cold-atom platform for spin-based sensing and open pathways toward quantum-enhanced searches for physics beyond the Standard Model.
}



\maketitle

\section*{Introduction}

Atomic-spin-based sensors are indispensable in quantum information science \cite{Luca2018QMetrologyRev} and precision measurement \cite{Budker2007Mag, Lei2025SpinExoticRMP}. Paramagnetic atoms, whose spins carry Bohr magnetons, underpin some of the most sensitive magnetometers \cite{Kominis2003SERF}. In contrast, diamagnetic atoms possess spins originating solely from their nuclei. These nuclear spins, governed by much smaller nuclear magnetons, exhibit significantly longer coherence times \cite{Zheng2022YbEDM}, making them ideal for sensing interactions beyond magnetism. Nuclear magnetic resonance (NMR) is a prominent example of such sensing.

In order to suppress magnetic-field noise, a comagnetometer employs two atomic species with distinct gyromagnetic ratios, colocated and measured simultaneously \cite{Terrano2022ComagNewPhys}. The ratio of their Larmor frequencies —termed the comagnetometer ratio—is intrinsically insensitive to magnetic-field fluctuations while retaining sensitivity to nonmagnetic interactions. Such systems have demonstrated the exceptional ability to detect energy shifts at the pHz level \cite{Vasilakis2009ComagSpinDependent}.

Room-temperature gas-cell comagnetometers have been extensively developed, with combinations including the $^{85}\mathrm{Rb}$-$^{87}\mathrm{Rb}$ paramagnetic pair \cite{Kimball2013ComagSpinGravity}, paramagnetic-diamagnetic pairs $\mathrm K$-$^3\mathrm{He}$ \cite{Vasilakis2009ComagSpinDependent} and $\mathrm {Rb}$-$^{21}\mathrm{Ne}$ \cite{Almasi2020ComagSpinSpin}, and diamagnetic pairs $^3\mathrm{He}$-$^{129}\mathrm{Xe}$ \cite{Cane2004ComagLorentz&CPT, Allmendinger2014ComagLorentz&CPT, Almasi2020ComagSpinSpin}, 
Xe isotopes \cite{Bulatowicz2013ComagAxionlike, Feng2022XeComagM-N} and Hg isotopes \cite{Lamoreaux1986ComagLorentz}. These platforms support a wide range of applications, from inertial sensing \cite{Kornack2005ComagGyroscope} to precision tests of fundamental physics, including searches for electric dipole moments (EDMs) \cite{Rosenberry2001ComagEDM, Sachdeva2019ComagEDM, Abel2020ComagnEDM}, monopole–dipole interactions \cite{Bulatowicz2013ComagAxionlike, Tullney2013SpinDependent, Feng2022XeComagM-N}, spin–gravity couplings \cite{Venema1992HgComagSpinGravity, Vasilakis2009ComagSpinDependent, Kimball2013ComagSpinGravity, Zhang2023ComagSpinGravity}, and axion-like dark matter \cite{Wu2019ComagDarkMatter, Alonso2019ComagDarkMatter}. However, their centimeter-scale cell sizes and millimeter-thick walls impose a limitation on sensitivity to short-range interactions with ranges below $\sim100\ \mu$m.

Cold-atom comagnetometers can overcome this limitation. Atom clouds confined in a laser trap can be reduced to scales of 1–100 $\mu$m, enabling sensitivity to comparable interaction ranges. Moreover, the absence of material walls eliminates related systematic effects. A prior implementation used two hyperfine states of paramagnetic $^{87}$Rb  with spin coherence times at several milliseconds \cite{Pau2020BECComag}. Signals of the two hyperfine states were combined to generate a single-species magnetometer based on the nuclear spin with a coherence time of 1 s. More recently, a cold-atom system combining paramagnetic $^{87}$Rb with diamagnetic $^{87}$Sr was realized \cite{Thekkeppatt2025ComaggFactor}. Operating under magnetic fields of 100-1000 G, this system enabled ppm-level determination of the nuclear $\mathit g$ factor of $^{87}$Sr, though its coherence time was ultimately limited to milliseconds by the $^{87}$Rb component under strong magnetic fields.


Here, we realize a comagnetometer based purely on nuclear spins consisting of $^{171}$Yb (spin-1/2) and $^{173}$Yb (spin-5/2), simultaneously confined in a one-dimensional optical lattice under a weak magnetic field (10-100 mG). A key challenge in such systems is the suppression of lattice-induced light shifts. We eliminate vector light shifts by enforcing linear polarization of the lattice beam. Tensor shifts are naturally absent in $^{171}$Yb due to its spin-1/2 character. For the higher-spin $^{173}$Yb, we apply quantum control techniques to prepare a Schr\"{o}dinger cat state, whose precession frequency is rendered insensitive to tensor shifts. Using simultaneous Ramsey interferometry on both isotopes, we measure the comagnetometer ratio and demonstrate a spin coherence time of 60 s. We identify a novel systematic effect in the cat state stemming from the interplay between Zeeman and tensor light shifts. By understanding and controlling both vector and tensor light shifts, we achieve a magnetic-noise suppression factor exceeding $3\times10^4$. We determine the ratio of nuclear magnetic moments between $^{171}$Yb and $^{173}$Yb with a precision of 4 ppm.

\section*{Realization of cold-atom comagnetometer}

\begin{figure}[htbp]
    \centering
    \includegraphics[width=1\textwidth]{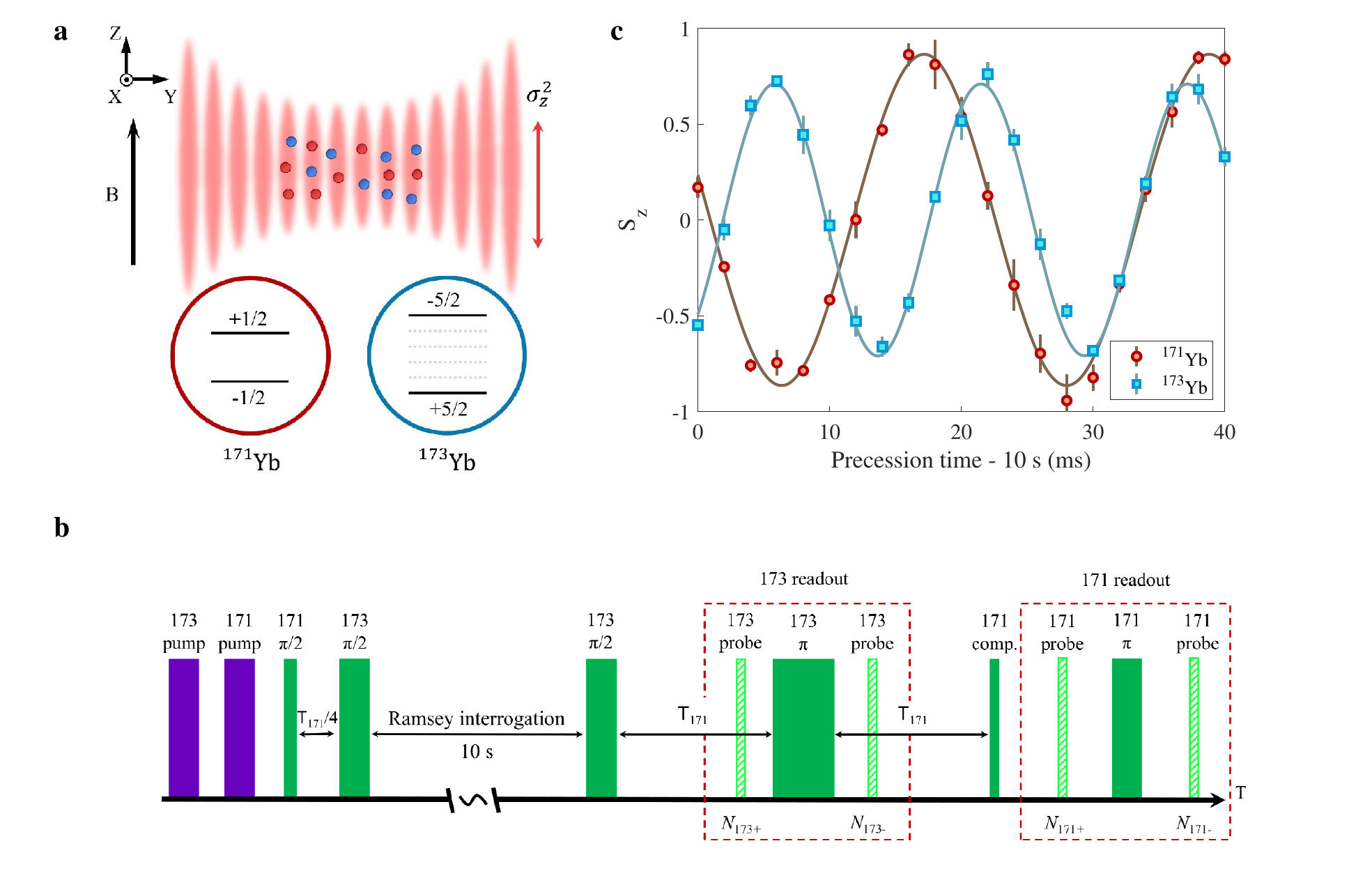}
    \caption{
     $^{171}$Yb-$^{173}$Yb cold-atom comagnetometer. 
     (a) Atoms of $^{171}$Yb and $^{173}$Yb are mixed and confined together in the optical lattice. A uniform field of 10-100 mG is applied to define the quantization axis in the vertical Z-direction. While $^{171}$Yb (I = 1/2) naturally has no tensor light shifts, $^{173}$Yb (I = 5/2) is prepared into the cat state, which is immune to decoherence caused by the tensor shifts in the optical lattice.
     (b) Measurement sequence of the simultaneous Ramsey interferometry. The arrangement of the 399 nm pump pulses (solid purple), 556 nm Rabi pulses (solid dark green), and 556 nm probe pulses (hatched light green) are explained in details in the main text.
     (c) Ramsey signals at 10 s with a bias field at 50 mG, where red circles represent the precession signals of $^{171}$Yb, and blue squares correspond to those of $^{173}$Yb. Error bars: one standard deviation (1-$\sigma$) derived from 3 repeated measurements; solid curves: least-squares fittings with sinusoidal functions.
     }
    \label{fig1}
\end{figure}

In preparation of a cold-atom ensemble, ytterbium atoms containing both $^{171}$Yb and $^{173}$Yb (isotopic abundances: 14\% and 16\%) exit an oven, pass through a Zeeman slower, and are captured by a blue magneto-optical trap (MOT) formed with a laser exciting the broad transition ${6s^2}\ {^{1}\mathrm{S}_0}\rightarrow{6s6p\ ^{1}\mathrm{P}_1}$ at 399 nm. The atoms are subsequently cooled to a temperature of $~\sim20\ \mu$K with a green MOT at 556 nm acting on the narrow transition ${6s^2}\ {^{1}\mathrm{S}_0}\rightarrow{6s6p\ ^{3}\mathrm{P}_1}$, and then transferred to a movable optical dipole trap (Bus ODT) at 1036.4 nm, which is a magic wavelength for the cooling transition \cite{Zheng2020Magic}. Carried by the Bus ODT from the MOT chamber to an adjacent science chamber across 0.35 m, the atoms arrive at their final destination: a horizontally directed, one-dimensional optical lattice also at 1036.4 nm (Fig.~\ref{fig1}a). With a beam power of 30 W and a waist of 25 $\mu$m, the lattice consists of trap sites with a depth of $\sim$ 2 mK. A total of about $2\times10^4$ $^{171}$Yb atoms and $2\times10^4$ $^{173}$Yb atoms are mixed together to form an ellipsoidal cloud with a diameter of 40 $\mu$m and a length of 80 $\mu$m, occupying approximately 160 lattice sites. The lifetime of atoms in the optical lattice is observed to be 50 s.

The science chamber is surrounded by a set of coils and magnetic shields. A uniform magnetic field of 10 - 100 mG in the Z-direction is applied to define the spin quantization axis (Fig.~\ref{fig1}a). In the $^{1}\mathrm S_0,\mathit F=\mathrm 1/2$ ground state of $^{171}$Yb, a transversely polarized state $\left\vert\psi_{t}\right\rangle = \left\vert F=1/2,m_F=+1/2\right\rangle_g+\mathrm{e}^{\mathrm{i}\phi_1}\left\vert\mathit F=\mathrm 1/2,\mathit m_F=\mathrm-1/2\right\rangle_{\mathit g}$ is employed to sense the precession phase $\phi_{1}$. For the ground state of $^{173}$Yb  (I = 5/2), it is necessary to prepare atoms into the cat state, $\left\vert\psi_{c}\right\rangle = \left\vert5/2,+5/2\right\rangle_g+\mathrm{e}^{\mathrm{i}\phi_2}\left\vert5/2,-5/2\right\rangle_g$, in order to achieve immunity to otherwise significant decoherence caused by tensor light shifts in the optical lattice \cite{Yang2025Cat}. The cat state cannot simply be prepared with either optical pumping or a rotating magnetic field. Instead, it is generated using a 556 nm Rabi beam, a circularly polarized laser beam traveling in the X-direction to drive the necessary Rabi transitions, with its $\sim$ GHz detunings from the ${6s^2}\ {^{1}\mathrm{S}_0}\rightarrow{6s6p\ ^{3}\mathrm{P}_1}$ transitions designed to induce a specific combination of vector and tensor light shifts \cite{Yang2025Cat}. For convenience, the same Rabi beam is also used to drive Rabi transitions in $^{171}$Yb, where it induces a pure vector shift in the spin-1/2 system. In this work, the induced Rabi transition frequency is $2\pi\times0.4$ kHz for $^{173}$Yb, and $2\pi\times3$ kHz for $^{171}$Yb.

Ramsey interferometry is simultaneously implemented on both isotopes. Its measurement sequence is illustrated in Fig.~\ref{fig1}b. Since all laser pulses are simultaneously shone on atoms of both isotopes, care needs to be taken either to avoid crosstalks between the two isotopes, or to compensate for any crosstalk effects. First, optical pumping at 399 nm along the Z-axis is applied to polarize $^{173}$Yb and $^{171}$Yb successively. Then, a $\pi/2$-pulse of the 556 nm Rabi beam is applied to
drive the spin state of $^{171}$Yb. The pulse also induces a small rotation ($\sim\pi/16$) on the spin of $^{173}$Yb. After a quarter of the $^{171}$Yb-Larmor period ($\mathrm T_{171}/4$), a $\pi/2$-pulse of the Rabi beam is applied to drive $^{173}$Yb into the cat state, while having little effect on $^{171}$Yb because their spins are pointing along the Rabi beam direction at the time of the pulse. Afterwards, spins of both isotopes are allowed to precess freely and accumulate phases over time. At the readout stage, first on $^{173}$Yb, the phase information is converted into population values by applying a $\pi/2$-pulse to the precessing cat state. The populations are measured as follows (see Methods): the first probe pulse at 556 nm measures the population of $\left\vert5/2, +5/2\right\rangle_g$, yielding $N_{173+}$; a subsequent $\pi$-pulse exchanges the populations between $\left\vert5/2,+5/2\right\rangle_g$ and $\left\vert5/2,-5/2\right\rangle_g$; the second probe then captures the original population of $\left\vert5/2,-5/2\right\rangle_g$, denoted as $N_{173-}$. The differential signal is derived as $S_{z} = (N_{+}-N_{-})/(N_{+}+N_{-})$. Note that both the second $\pi/2$-pulse and the $\pi$-pulse on $^{173}$Yb unavoidably operate on $^{171}$Yb as well. To compensate their effects on $^{171}$Yb, an additional pulse is introduced so that the three pulses, each successive one delayed by a complete $^{171}$Yb-Larmor period ($\mathrm T_{171}$), combine to rotate the spin of $^{171}$Yb by (12 + 1/2)$\pi$ . Finally, the populations of $^{171}$Yb spin states are read out in a way similar to that of $^{173}$Yb.

The resulting Ramsey fringes are displayed in Fig.~\ref{fig1}c. The contrast of $^{173}$Yb decays due to interspecies collision between $^{173}$Yb and $^{171}$Yb, reducing the coherence time of $^{173}$Yb from thousands of seconds down to 60 s. Meanwhile, the collisions seem to have no degradation effects on $^{171}$Yb. 

By applying simultaneous Ramsey interferometry on both isotopes, we determine the precession phases, and derived the Larmor frequencies of $^{171}$Yb ($\omega_{171}$) as well as the cat state of $^{173}$Yb ($\omega_{173}$), both with a statistical sensitivity of $\sim\ 2\pi\times2$ mHz in a single-shot measurement.

Under ideal conditions, the comagnetometer frequency ratio $R_L = \omega_{171}/\omega_{173}$ is insensitive to magnetic-field fluctuation. However, we find that light shifts induced by the high-power optical lattice could cause a significant perturbation to $R_L$.

\section*{Dependence on light shift}



The light shift can be decomposed into scalar, vector, and tensor terms, which are respectively proportional to the zeroth, first, and second powers of the magnetic quantum number ${m_F}$ \cite{LeKien2013Polarizability}. The scalar shift causes identical displacements for all ${m_F}$ states, thereby contributing no phase shifts in the evolution of angular momentum states. The vector shift, induced by circularly polarized light, acts as a Zeeman-like shift with its quantization axis along the wave vector of the light field. Given the linearly polarized optical lattice used in this work, the tensor shift aligned along the polarization direction emerges as the predominant effect.

\subsection*{Tensor shift}

Tensor shifts are naturally absent in the ground state $^{1}\mathrm{S}_0$ of $^{171}$Yb (I = 1/2). For $^{173}$Yb (I = 5/2), tensor shifts are also expected to be null in the cat state of $^{1}\mathrm S_0$, but instead are found in this work to affect the Larmor frequency in a significant and peculiar way. The tensor operator aligns along the linear polarization direction (z) of the lattice, which, due to imperfection, form a small angle $\theta$ with the bias magnetic field $\mathrm B$ in the Z-direction. Without loss of generality, assume the magnetic field has a small component in the transverse x-direction, the Hamiltonian $\mathscr H$ arising from the combined magnetic and tensor fields can be expressed as

\begin{equation}\label{Ham}
    \mathscr H = \Omega_{\mathrm B}(\sigma_{z}+\theta\sigma_{x})+\Omega_{\mathrm T}\sigma_{z}^2\ (\theta\ll1).
\end{equation}
Here, $\sigma_{z}$ and $\sigma_{x}$ are the angular momentum matrices, with their terms corresponding to Zeeman effects. The $\sigma_{z}^2$ term describes the tensor effect. $\Omega_{\mathrm B}$ and $\Omega_{\mathrm T}$ are the corresponding coupling parameters.

\begin{figure}[htbp]
    \centering
    \includegraphics[width=1\textwidth]{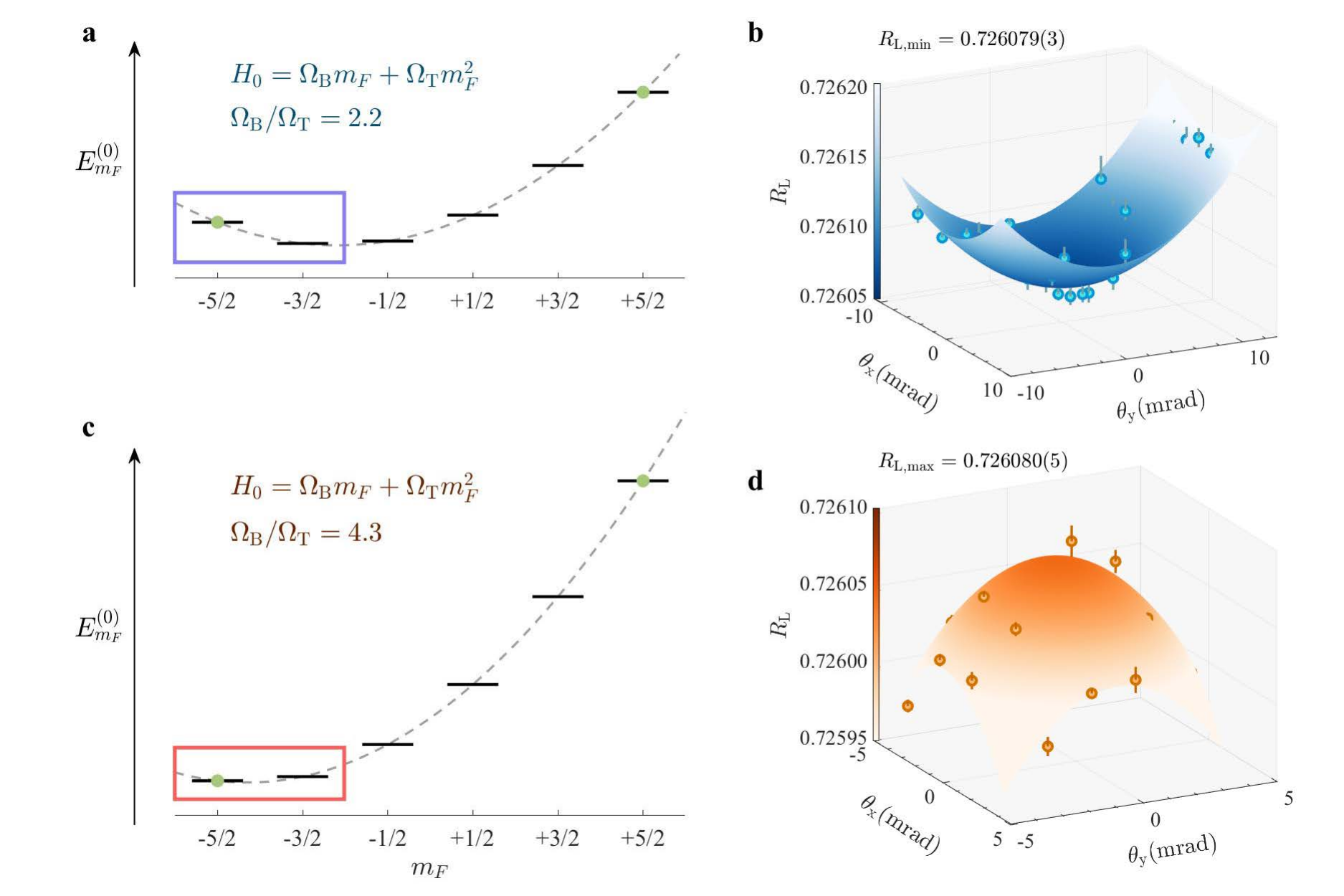}
    \caption{The combined effect of Zeeman and tensor shifts on the Larmor frequency ($\omega_{173}$) of the cat state of $^{173}$Yb, and on the comagnetometer ratio $R_L=\omega_{171}/\omega_{173}$. $^{173}$Yb atoms are populated in $\left\vert5/2,\pm5/2\right\rangle_g$ (green dots). 
    (a) At $\Omega_{\mathrm B}/\Omega_{\mathrm T}=2.2$, where $\left\vert5/2,-5/2\right\rangle_g$ is shifted higher than $\left\vert5/2,-3/2\right\rangle_g$, the $\sigma_x$ term mixes the two states in the rectangle and pushes $\left\vert5/2,-5/2\right\rangle_g$ further up. This results in a lower frequency $\omega_{173}$, and a higher $R_L$ value.
    (b) Under 50 mG, corresponding to $\Omega_{\mathrm B}/\Omega_{\mathrm T}=2.2$, $R_L$ is mapped as the magnetic-field orientation is scanned around the Z-direction. The higher $R_L$ values in (a) results in a surface with positive curvatures.
    (c) At $\Omega_{\mathrm B}/\Omega_{\mathrm T}=4.3$, where $\left\vert5/2,-5/2\right\rangle_g$ is shifted below $\left\vert5/2,-3/2\right\rangle_g$, the $\sigma_x$ term pushes $\left\vert5/2,-5/2\right\rangle_g$ further down.
    (d) Under 100 mG, corresponding to $\Omega_{\mathrm B}/\Omega_{\mathrm T}=4.3$. The curvature of the surface is found to reverse.
    Error bars: one standard deviation (1-$\sigma$) derived from 50 repeated measurements; surfaces: least-squares fittings with quadratic surface functions.
    }
    \label{fig2}
\end{figure}

Under constant optical trapping conditions, we scan the orientation of the bias magnetic field over two dimensions and map the comagnetometer ratio $R_L$. First, the magnetic field is set at 50 mG, corresponding to $\Omega_{\mathrm B}/\Omega_{\mathrm T}=2.2$ (see Methods). The measured $R_L$ values fall onto a parabolic surface with a minimum extremum $R_{L,min} = 0.726079(3)$ (Fig.~\ref{fig2}b). Next, at a higher magnetic field of 100 mG, where $\Omega_{\mathrm B}/\Omega_{\mathrm T}=4.3$, the opening direction of the parabolic surface is found to be inverted, yielding a maximum $R_{L,max} = 0.726080(5)$ (Fig.~\ref{fig2}d), with its value consistent with the minimum value obtained at 50 mG. The parabolic surface and its curvature inversion can be attributed to a second-order perturbation effect on the $^{173}$Yb energy levels, with the transverse $\sigma_{x}$ term treated as a perturbation. The resulting shifts on the energy splitting between $\left\vert5/2,+5/2\right\rangle_g$ and $\left\vert5/2,-5/2\right\rangle_g$ can be expressed as 

\begin{equation}\label{perturbation2nd}
\begin{aligned}
    \Delta E := \delta E_{+5/2}^{(2)}-\delta E_{-5/2}^{(2)} 
    = \dfrac{5}{2} \cdot \left(\frac{\Omega_{\mathrm B}}{\Omega_{\mathrm T}}\right)^2 \cdot \dfrac{ \Omega_{\mathrm B} \theta^2}  {(\frac{\Omega_{\mathrm B}}{\Omega_{\mathrm T}} + 4)(\frac{\Omega_{\mathrm B}}{\Omega_{\mathrm T}} - 4)} ,
\end{aligned}
\end{equation}
where $\delta E_{m_F}^{(2)}$ denotes the $2^{nd}$-order perturbation correction to the energy level of $\left\vert5/2,m_F\right\rangle_{g}$. Derivation of Eq.~(\ref{perturbation2nd}) is presented in Methods. It can be seen that, on opposite sides of the critical ratio $\Omega_{\mathrm B}/\Omega_{\mathrm T}=4$, the sign of $\Delta E$ reverses, resulting in
curvature inversion of the $R_L$ surface. At the critical ratio of $\Omega_{\mathrm B}/\Omega_{\mathrm T}=4$, due to the degeneracy between $\left\vert5/2,-5/2\right\rangle_g$ and $\left\vert5/2,-3/2\right\rangle_g$, the above perturbation treatment is no longer valid, and the cat state can no longer be preserved.

In order to evaluate the robustness of the comagnetometer against environmental magnetic field fluctuations, we define the suppression factor $\eta$ as:

\begin{equation}
\eta({\mathrm B})=\left\lvert \dfrac{\Delta {\mathrm B}}{\mathrm B} \Big/ \dfrac{\Delta R_L}{R_L} \right\rvert.
\end{equation}
From 50 mG to 100 mG, the magnetic field changes by a factor of two, so are the Larmor frequencies of each isotope, yet the comagnetometer ratio $R_L$ remains constant at the 10 ppm level. We derive the suppression factor $\eta>3\times10^4$ ($95\%$ confidence level) based on the two $R_L$ values and their corresponding uncertainties. 

\subsection*{Residual vector shift}

In order to minimize vector shifts, the optical lattice is tuned to be as close to linearly polarized as possible. This is done by optimizing the coherence of the Ramsey interferometry of $^{171}$Yb. During polarization tuning, the bias magnetic field is temporarily redirected to the horizontal propagation direction of the optical lattice (Y-axis), and its strength lowered to 1 mG, both measures taken to deliberately enhance the decoherence effect due to the inhomogeneous vector shifts in the lattice. In this way, the residual circular polarization component, as measured by the Stokes parameter $S_3$, is tuned to be less than 0.1\%. According to calculated vector polarizabilities of $^{171}$Yb and $^{173}$Yb, the vector shifts due to the residue circular polarization can be regarded as effective magnetic fields less than 100 $\mu$G along the Y-direction. At a bias field of 50 mG in the Z-direction, the relative deviation in $R_L$ due to the residue vector shift is estimated to be less than $3\times10^{-7}$, which is negligible compared to the current measurement uncertainty of $3\times10^{-6}$. When the field is increased to 100 mG, the deviation drops further down to $8\times10^{-8}$.

\section*{Ratio of nuclear magnetic moments}

\begin{figure}[htbp]
    \centering
    \includegraphics[width=1\textwidth]{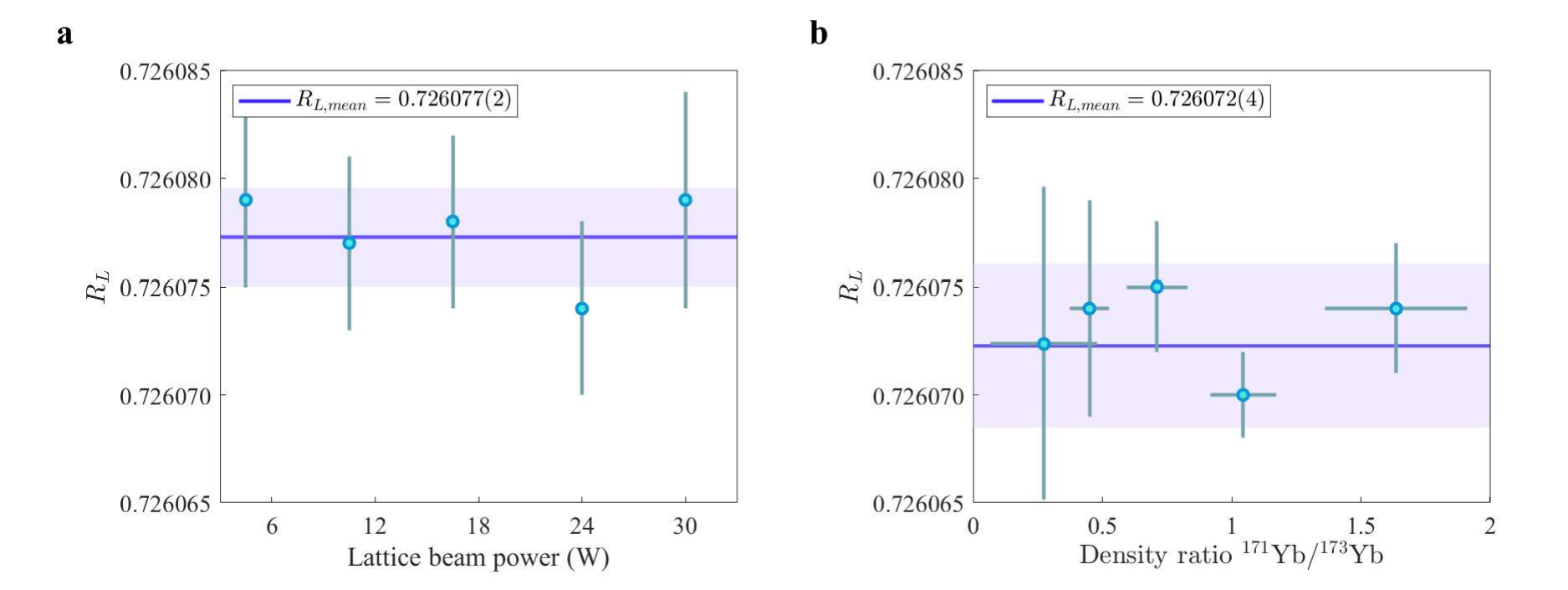}
    \caption{Investigating possible systematic effects.
    (a) In a magnetic field of 100 mG, the lattice beam power is varied.
    (b) With both the magnetic field and lattice parameters fixed, the density ratio between $^{171}$Yb and $^{173}$Yb is varied. No dependences on these experimental parameters are observed.
    Error bars: one standard deviation (1-$\sigma$) derived from 50 repeated measurements; solid lines: weighted averages.
    }
    \label{fig3}
\end{figure}

We have investigated systematic uncertainties in the measurements of the comagnetometer ratio $R_L$. After aligning the 100 mG magnetic field to be parallel to the polarization direction of the optical lattice, we measure $R_L$ while varying the lattice power by more than a factor of five. No dependence of $R_L$ on light power is observed within the measurement uncertainties (Fig.~\ref{fig3}a). A weighted average across different lattice powers yields $R_L = 0.726077(2)$.
We measure $R_L$ while varying the density ratio of $^{171}$Yb over $^{173}$Yb by a factor of seven. This is to investigate possible collisional shifts that are likely to be density dependent. Again, no dependence of $R_L$ on the density ratio is observed (Fig.~\ref{fig3}b). The average value across different density ratios is 0.726072(4). The Earth’s rotation frequency of $2\pi\times11.6\ \mu\mathrm{Hz}$ is negligible compared to the measurement uncertainty of around $2\pi\times100\ \mu\mathrm{Hz}$ for the Larmor frequencies.

We calculate a weighted average of comagnetometer ratio  $R_L$ across different experimental parameters, yielding a final value of $R_L = 0.726076(3)$ and the ratio of nuclear magnetic moments between $^{171}$Yb and $^{173}$Yb, $R_{NMM} = \mu_{I,171}/\mu_{I,173} = -0.726076(3)$. This new result agrees with the recommended value of $-0.72610(4)$ \cite{Gossard1964RefNMR, Olschewski1972Ref, Stone2005Ref}, while improving the precision by an order of magnitude.

\section*{Discussion}

With its strong suppression of magnetic-field noise and tight atom localization within the lattice sites, our cold-atom comagnetometer enables sensitive searches for exotic interactions beyond the Standard Model, particularly at short interaction ranges of 1-100 $\mu$m. Among many new-physics interactions proposed \cite{Safronova2018NewPhys, Lei2025SpinExoticRMP}, we present two examples highlighting the capability of this system.


The comagnetometer is used to search for $\mathit P$- and $\mathit T$-violating spin-mass couplings, also known as monopole-dipole interactions \cite{Moody1984M-N, Bulatowicz2013ComagAxionlike, Tullney2013SpinDependent, Feng2022XeComagM-N}. In this search, with $^{171}$Yb and $^{173}$Yb atoms as spin sources, the comagnetometer ratio is monitored as a dense mass source is brought into proximity with the atoms. By using optical tweezers to position the atoms, the atom-mass separation can be precisely controlled down to a few micrometers. We estimate that this setup can improve the sensitivity to spin-mass couplings at the 1 $\mu$m interaction range by up to three orders of magnitude compared to the current experimental limits \cite{Guigue2015Spin3HePRD}.


We aim to improve the search sensitivity of electric dipole moments (EDMs) using cold atoms \cite{Bishof2016RaEDM, Zheng2022YbEDM}. The current EDM experiments on $^{171}$Yb \cite{Zheng2022YbEDM} are limited by magnetic-field noises, particularly the Johnson noise in adjacent electrodes. A $^{171}$Yb-$^{173}$Yb cold-atom comagnetometer can effectively suppress such noise effects, thereby boosting the EDM search sensitivity. Furthermore, EDM searches have been performed on cold atoms of $^{225}$Ra (I = 1/2), an isotope whose octupole-deformed nucleus provides enhanced sensitivity to $\mathit T$-violating interactions \cite{Bishof2016RaEDM}. As a future direction, we propose forming a comagnetometer using $^{223}$Ra (I = 3/2) and $^{225}$Ra to suppress magnetic noise in the next-generation $^{225}$Ra EDM experiments.


\section*{Methods}
\subsection* {Zeeman and tensor light shifts of $^{173}$Yb}\label{CombinedShift}


The Hamiltonian in Eq.~(\ref{Ham}) is rewritten below, treating the $\sigma_x$ term as a perturbation:

\begin{equation}\label{Ham_pur}
\begin{aligned}
    \mathscr H & = \mathscr{H}_0+\theta\mathscr{H}' \\
    & = \Omega_{\mathrm B}\sigma_z + \Omega_{\mathrm T}\sigma_z^2 + \theta\Omega_{\mathrm B}\sigma_x\ (\theta\ll 1).
\end{aligned}
\end{equation}

The perturbated eigenenergies are

\begin{equation}\label{perturbation}
    E_{m_F} = E_{m_F}^{(0)}
     +\theta\left\langle \psi_{m_F}^{(0)}\right\vert\mathscr{H}'\left\vert\psi_{m_F}^{(0)}\right\rangle
    +\sum_{i\neq {m_F}}\theta^2    \dfrac{{\left\langle\psi_i^{(0)}\right\vert\mathscr{H}'\left\vert\psi_{m_F}^{(0)} \right\rangle}^2}
    {E_{m_F}^{(0)}-E_i^{(0)}}
    +O(\theta^3),
\end{equation}

\begin{equation}\label{E(0)}
    E_{m_F}^{(0)}=m_F\Omega_{\mathrm B}+m_F^2\Omega_{\mathrm T},
\end{equation}
where the $\theta^2$ term corresponds to $\delta E_{m_F}^{(2)}$ in Eq.~(\ref{perturbation2nd}). When $\Omega_{\mathrm B}/\Omega_{\mathrm T}=4$, the unperturbed energy levels of $\left\vert5/2,-5/2\right\rangle_g$ and $\left\vert5/2,-3/2\right\rangle_g$ are degenerate.

Here the first-order perturbation term is always zero because the $\sigma_x$ term corresponds to a subdiagonal matrix. 
For the spin-5/2 system in $^{173}$Yb, the $\sigma_x$ matrix is

\begin{equation}
\sigma_x = 
    \begin{pmatrix}
        0 & \frac{\sqrt{5}}{2} & & & & \\
        \frac{\sqrt{5}}{2} & 0 & \frac{\sqrt{8}}{2} & & & \\
         & \frac{\sqrt{8}}{2} & 0 & \frac{\sqrt{9}}{2} & & \\
         & & \frac{\sqrt{9}}{2} & 0 & \frac{\sqrt{8}}{2} & \\
         & & & \frac{\sqrt{8}}{2} & 0 & \frac{\sqrt{5}}{2}\\
         & & & & \frac{\sqrt{5}}{2} & 0 \\
    \end{pmatrix}.
\end{equation}

To calculate the Larmor frequency of the cat state of $^{173}$Yb, we consider the $2^{nd}$-order perturbation on the two populated stretched states $\left\vert5/2,\pm5/2\right\rangle_g$:

\begin{equation}\label{E+5/2}
    \delta E_{+5/2}^{(2)} = \theta^2\cdot\dfrac{5}{4}\cdot\dfrac{\Omega_{\mathrm B}^2}{\Omega_{\mathrm B}+4\Omega_{\mathrm T}},
\end{equation}

\begin{equation}\label{E-5/2}
    \delta E_{-5/2}^{(2)} = \theta^2\cdot\dfrac{5}{4}\cdot\dfrac{\Omega_{\mathrm B}^2}{-\Omega_{\mathrm B}+4\Omega_{\mathrm T}}.
\end{equation}
Subtracting the two energy shifts yields Eq.~(\ref{perturbation2nd}).


\subsection*{Calibration of $\Omega_{\mathrm B}/\Omega_{\mathrm T}$ }

We perform precession measurements on $^{173}$Yb to calibrate the $\Omega_{\mathrm B}/\Omega_{\mathrm T}$ values. Since the cat state is immune to tensor shifts ($\Omega_{\mathrm T}$), we employ a coherent spin state (CSS) instead, whose spin is fully polarized along a direction.

\begin{figure}[htbp]
    \centering
    \includegraphics[width=0.7\textwidth]{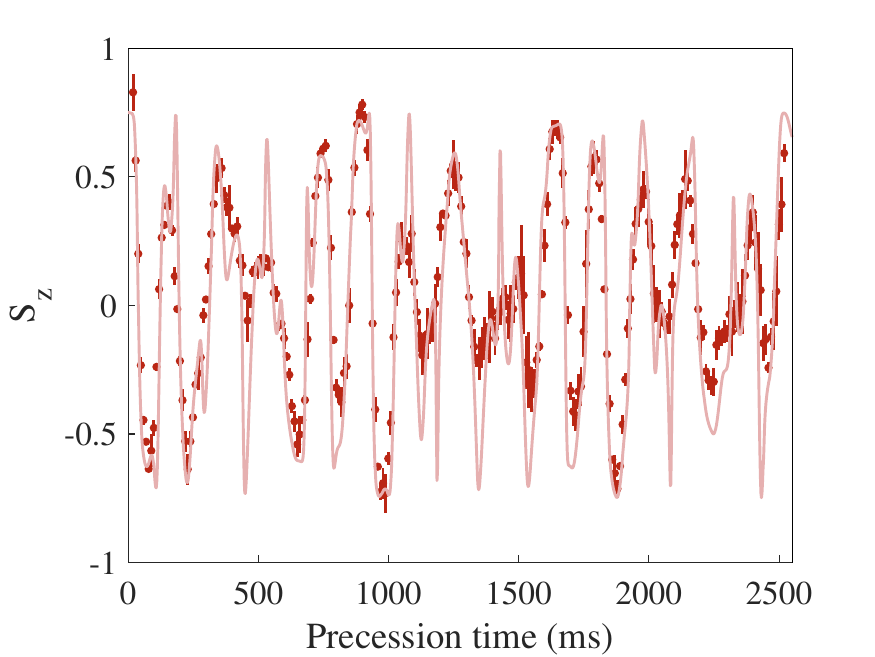}
    \caption {Precession of a coherent spin state of $^{173}$Yb under $\mathrm B_{\mathit x}$ = 23.5 mG and 8 W lattice power. Error bars: one standard deviation (1-$\sigma$) derived from 3 repeated measurements. The solid curve represents simulation results.}
     \label{fig4}
\end{figure}

While keeping the lattice linearly polarized along the z-axis, its power is reduced to 8 W. Meanwhile, a bias magnetic field of 23.5 mG is applied along the x-direction. The $^{173}$Yb atoms are first pumped into $\left\vert5/2,+5/2\right\rangle_g$, then allowed to precess under the influence of both the Zeeman and tensor shifts, $\mathscr{H}_{calib} = \Omega_{\mathrm B}\sigma_x + \Omega_{\mathrm T}\sigma_z^2$. After a variable precession duration (0 – 2,500 ms), the population signal $S_z$ is measured along the z-axis (Fig.~\ref{fig4}). 



We simulate the precession process of $^{173}$Yb using the calibration Hamiltonian ($\mathscr{H}_{calib}$) and compare the results with experimental data, obtaining $\Omega_{\mathrm B}/\Omega_{\mathrm T}=3.5$ under the given magnetic field and lattice intensity conditions.


\subsection*{Atom sample detection}

\begin{figure}[htbp]
    \centering
    \includegraphics[width=1\textwidth]{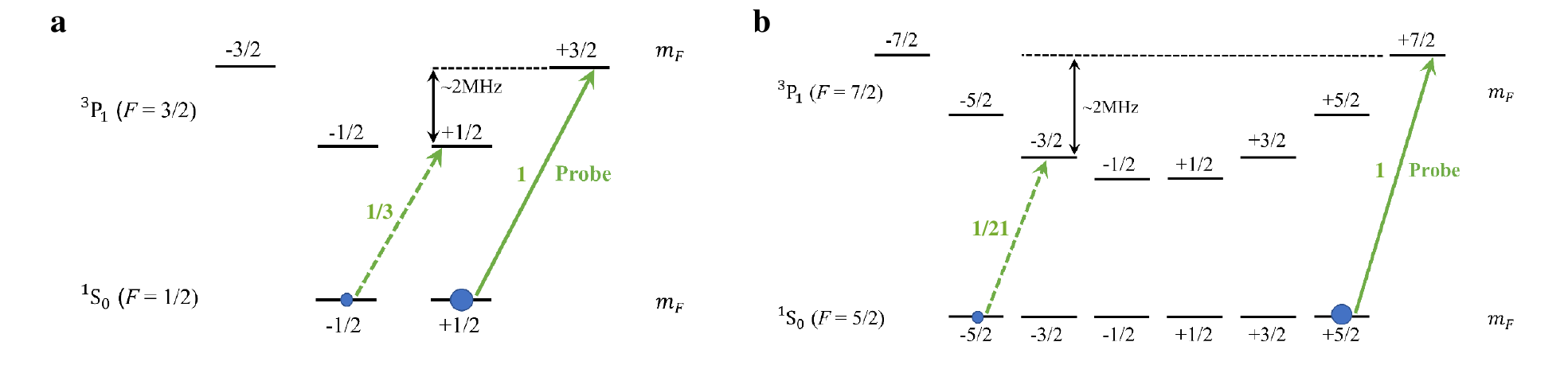}
    \caption{
     Transitions used to measure spin-state populations in $^{171}$Yb (a) and $^{173}$Yb (b). The arrow of solid line indicates a cycling transition used for population measurements; the arrow of dashed line indicates a spin-flip transition suppressed by large detunings.}
     \label{fig5}
\end{figure}

A circularly polarized laser probe beam along the Z-direction runs through the cold atom ensemble and is detected by a CMOS camera. Absorption images are taken on the atoms for population measurements. The state readout of atoms is performed via differential detection of two stretched states. For $^{171}$Yb atoms, the probe beam resonantly excites the cycling transition $\left\vert1/2,+1/2\right\rangle_g \rightarrow \left\vert{3/2},+3/2\right\rangle_e$ (Fig.~\ref{fig5}a). Meanwhile, the $\left\vert1/2,-1/2\right\rangle_g \rightarrow \left\vert3/2,+1/2\right\rangle_e$ transition is off resonance by 10 times the natural linewidth ($10\ \Gamma$) due to tensor shifts on the excited states. As a result, the cycling transition can be repeated 1,200 times on average before a spin flip occurs. This spin selectivity ensures that the spin state is detected with an efficiency approaching $100\%$. For $^{173}$Yb atoms, the relevant transitions are $\left\vert5/2,+5/2\right\rangle_g \rightarrow \left\vert7/2,+7/2\right\rangle_e$ and $\left\vert5/2,-5/2\right\rangle_g \rightarrow \left\vert3/2,-3/2\right\rangle_e$ (Fig.~\ref{fig5}b). In addition to a comparable difference of tensor shifts ($\sim20\ \Gamma$), the branching ratios of these two transitions differ by a factor of 21, thus further enhancing the spin selectivity.

\section*{Acknowledgements}
We thank D. Sheng for fruitful discussions and critical reading of the manuscript, and thank S.-Z. Wang for contributions to the apparatus during the early stages. 

\section*{Declarations}


\begin{itemize}
\item Funding

This work is supported by the National Natural Science Foundation of China (NSFC) through grant no. 12174371, the Innovation Program for Quantum Science and Technology through grant no. 2021ZD0303101, the Strategic Priority Research Program of the Chinese Academy of Sciences through grant no. XDB21010200, and the Major Frontier Research Project of the University of Science and Technology of China through grant no. LS9990000002.

\item Author contribution

J.-L. Z. and W.-T. L. constructed the experimental apparatus and performed the experiments and simulations. Y. A. Y. contributed to the apparatus during the early stages. Y.-Q. W. made contributions to the theory part. J.-L. Z., W.-T. L., Y.-Q. W., T. X. and Z.-T. L. carried out data analysis and wrote the manuscript.

\end{itemize}



\end{document}